\documentclass[sigconf]{acmart}

\renewcommand\footnotetextcopyrightpermission[1]{} 
\renewcommand\footnotetextcopyrightpermission[1]{}
\renewcommand{\shortauthors}{}
\acmConference[ ]{}{}{}
\acmYear{}
\acmDOI{}
\acmISBN{}

\AtBeginDocument{%
  }

\setcopyright{acmlicensed}

\usepackage[normalem]{ulem}  
\usepackage{multirow}
\usepackage{graphicx}
\usepackage{booktabs, multirow, threeparttable}
\usepackage{makecell} 
\usepackage{pgfplots}
\usepgfplotslibrary{fillbetween} 
\usepackage{bm}

\usepgfplotslibrary{groupplots}
\usepackage{tablefootnote}

\usepackage{pgfplotstable}

\usepackage{pgfplots}
\usepackage{xcolor}
\usepackage{subcaption}
\usepackage[abs]{overpic} 

\usepackage{tabularx}
\usepackage{pifont}   

\usetikzlibrary{intersections,fillbetween}
\usetikzlibrary{patterns}
\begin{document}
\usetikzlibrary{intersections,fillbetween}

\title{XNet: A Sustainable System for Predicting Cross-Platform Narrative Diffusion via Discourse-Derived Communities}
\title{BRIDGE: A Unified Discourse Network\\for Platform-Invariant Narrative Diffusion Prediction}
\title{Cross-Platform Narrative Prediction:\\ Leveraging Platform-Invariant Discourse Networks}

\author{Patrick Gerard}
\email{pgerard@isi.edu}
\affiliation{%
  \institution{Information Sciences Institute}
  \city{Los Angeles}
  \state{CA}
  \country{USA}
}
\author{Luca Luceri}
\email{lluceri@isi.edu}
\affiliation{%
  \institution{Information Sciences Institute}
  \city{Los Angeles}
  \state{CA}
  \country{USA}
}
\author{Leonardo Blas}
\email{blasurru@usc.edu}
\affiliation{%
  \institution{University of Southern California}
  \city{Los Angeles}
  \state{CA}
  \country{USA}
}
\author{Emilio Ferrara}
\email{emiliofe@usc.edu}
\affiliation{%
  \institution{University of Southern California}
  \city{Los Angeles}
  \state{CA}
  \country{USA}
}
\renewcommand{\shortauthors}{Gerard et al.}

\begin{abstract}

Online narratives spread unevenly across platforms, with content emerging on one site often appearing on others, hours, days or weeks later. Existing cross-platform information diffusion models often treat platforms as isolated systems, disregarding cross-platform activity that might make these patterns more predictable. In this work, we frame cross-platform prediction as a network proximity problem: rather than tracking individual users across platforms or relying on brittle signals like shared URLs or hashtags, we construct platform-invariant discourse networks that link users through shared narrative engagement. 
We show that cross-platform neighbor proximity provides a strong predictive signal: adoption patterns follow discourse network structure even without direct cross-platform influence. 
Our highly-scalable approach substantially outperforms diffusion models and other baselines while requiring less than 3\%
of active users to make predictions. 
We also validate our framework through retrospective deployment.
We sequentially process a datastream of 5.7M social media posts occurred during the 2024 U.S. election, to simulate real-time collection from four platforms (X, TikTok, Truth Social, and Telegram):
our framework successfully identified emerging narratives, including crises-related rumors, yielding over 94\% AUC with sufficient lead time to support proactive intervention.


\end{abstract}






\maketitle


\pgfplotsset{
    colormap={cool}{
        rgb255(0cm)=(255,255,255);  
        rgb255(1cm)=(0,117,242);    
        rgb255(2cm)=(76,181,174)    
    }
}
\pgfplotsset{
    colormap={violet1}{color=(white) rgb255=(238,140,238) rgb255=(25,25,122)}
}

\definecolor{NoBridge}{HTML}{0075F2}        
\definecolor{Bridge}{HTML}{4CB5AE}          
\definecolor{co-url}{HTML}{b8b8ff}
\definecolor{co-hashtag}{HTML}{ffa9a3}
\definecolor{co-retweet}{HTML}{ffac81}
\definecolor{co-fast-retweet}{HTML}{ace1af}
\definecolor{kNN_Embedding_Graph}{HTML}{afcafc}
\definecolor{tkNN_Embedding_Graph}{HTML}{d3d3d3}
\definecolor{Embedding_Averaging}{HTML}{0075F2}
\definecolor{kNN_Embedding_Graph}{HTML}{4CB5AE}

\section{Introduction}

During the 2024 U.S. election, false narratives frequently spread across platform boundaries before reaching mainstream attention. The Springfield, Ohio, rumor that Haitan immigrants were eating pets spread from local Facebook groups through far-right networks on Gab and Telegram, later gaining traction on X (formerly Twitter) before ultimately being amplified at the Presidential Debate~\footnote{See \href{https://www.nbcnews.com/politics/donald-trump/trump-fringe-online-claim-immigrants-eating-pets-debate-trump-rcna170759}{NBC News report}}. Within days, the town was faced with over 30 bomb threats, forcing the evacuation of schools, hospitals, and government buildings~\footnote{See \href{https://www.nbcnews.com/news/us-news/30-bomb-threats-made-springfield-ohio-false-pets-claims-rcna171392}{NBC News report}}. Hurricane Helene FEMA conspiracies followed a similar trajectory: false claims that federal responders were blocking aid and seizing property emerged across platforms, prompting armed militia threats that forced emergency workers to evacuate and temporarily suspend relief efforts in affected areas~\footnote{See \href{https://www.theguardian.com/us-news/2024/oct/14/north-carolina-hurricane-helene-fema-armed-militia-threat}{The Guardian report}}. 


These patterns reveal a fundamental gap in our understanding of information flow across platforms. Content diffuses unevenly across platforms---narratives appearing on one site often precede subsequent appearance on others~\cite{ribeiro2021evolution,cinus2025exposing,minici2024uncovering}---yet existing methods tend to treat each platform as an isolated system~\cite{ng2022cross, ng2023coordinating}. We track hashtags on Twitter, forwarded messages on Telegram, and videos on TikTok as separate phenomena, missing the cross-platform connections that make diffusion patterns observable. This fragmentation has major real-world consequences. Coordinated influence campaigns exploit monitoring gaps by seeding narratives on loosely moderated platforms before mainstream amplification~\cite{minici2024uncovering, ng2022cross, zannettou2019web}, while platform API restrictions further fragment detection capabilities~\cite{tromble2021have}. 

We reframe this problem by treating the social media ecosystem as a unified information landscape. By constructing platform-invariant discourse networks~\cite{gerard2025bridging} that link users through shared narrative engagement, we exploit cross-platform homophily: users structurally close in this network should exhibit correlated adoption patterns~\cite{bollen2011happiness,mcpherson2001birds,barbera2015birds}, providing strong predictive signal for \textit{when} and \textit{where} narratives will emerge across platform boundaries.

On 5.7M posts from X, TikTok, Truth Social, and Telegram during the 2024 U.S. election, we show that representing users through platform-invariant discourse networks makes cross-platform diffusion predictable from neighbor activity alone. This reduction from complex multi-platform dynamics to neighbor activity patterns allows lightweight models to substantially outperform diffusion simulations and other baselines.
Applied retrospectively, this approach correctly identified high-impact narratives days before mainstream emergence, including the Hurricane Helene FEMA conspiracy, demonstrating potential for early warning in fragmented information ecosystems.

\textbf{Contributions.} Our main contributions include:
\begin{itemize}
\item \textbf{Platform-invariant proximity framework}: We reformulate cross-platform diffusion prediction from a complex multi-platform coordination modeling task to a neighbor activity analysis task via discourse network social proximity.
\item \textbf{Efficient prediction with minimal coverage}: Using only 2.9\% of active users, we substantially outperform diffusion simulations (+55\%) and next-best baselines (+27\%).

\item \textbf{Cross-platform narrative dataset}: We release 5.7M posts from X, TikTok, Truth Social, and Telegram with validated narrative labels and 2,943 cross-platform emergence events during the 2024 U.S. election.
\item \textbf{Operational validation}: Retrospective analysis showing the model would have correctly identified high-impact narratives (Hurricane Helene FEMA conspiracy, Springfield rumors) days before mainstream emergence.
\end{itemize}

\begin{table*}[t]
\centering
\caption{Cross-platform narrative variation: The same Hurricane Helene FEMA narrative manifests differently across platforms}
\label{tab:narrative_examples}
\renewcommand{\arraystretch}{1.3}
\small
\begin{tabularx}{\textwidth}{@{}l>{\raggedright\arraybackslash}X@{}}
\toprule
\small
\textbf{Platform} & \textbf{Expression of Narrative} \\
\midrule
\textbf{X/Twitter} & FEMA blocking private rescue ops in NC \#HurricaneHelene \#FEMA \\
\textbf{TikTok} & \textit{[Video transcript]} ``...apparently FEMA is telling volunteers they can't help hurricane victims? [inaudible] is actually happening here...'' \\
\textbf{Truth Social} & More evidence of federal incompetence: FEMA actively interfering with private relief efforts in North Carolina. \\
\textbf{Telegram} & FEMA interference with volunteer rescue operations confirmed by multiple sources. Patterns consistent with a deliberate resource control strategy. \\
\bottomrule
\end{tabularx}
\end{table*}

\section{Related Work}

\vspace{2pt} \noindent
\textbf{Network-Based User Representations.} A core challenge in cross-platform social media analysis is how to represent users in a way that captures their relationships, ideological alignment, and narrative engagement~\cite{barbera2015birds, gerard2025bridging, jiang2023retweet}. Traditional approaches build networks from platform-specific signals such as follower links~\cite{bollen2011happiness}, reposts and mentions~\cite{cinelli2021echo}, hashtags~\cite{alieva2022investigating, burghardt2024socio}, or URLs~\cite{nizzoli2021coordinated, luceri2024unmasking}. These signals depend on affordances that don't transfer (e.g., retweets on X vs. forwards on Telegram) and face increasing API restrictions~\cite{tromble2021have}.

Semantic approaches attempt to bypass these platform-based limitations by linking users based on content similarity~\cite{ng2023coordinating, luceri2024unmasking}. While promising, these methods often rely on expensive pairwise comparisons and remain sensitive to linguistic variation, limiting their scalability and robustness for early warning applications. Recent work on \textit{discourse networks} addresses these challenges by representing users through shared narrative participation rather than direct behavioral traces~\cite{gerard2025bridging}. This approach proves particularly valuable for cross-platform analysis, as narrative-level alignment can persist even when platform-specific behaviors differ.

\vspace{2pt} \noindent
\textbf{Cross-Platform Information Diffusion Prediction.} A growing body of work examines how narratives and events diffuse across online ecosystems, yet prediction remains methodologically challenging. Early approaches relied on heuristic user matching or link-based tracking, which are brittle and rarely scale across fragmented environments~\cite{iofciu2011identifying, cinus2024exposing}.  More recent work uses semantic similarity, clustering, or temporal modeling~\cite{hajiakhoond2019lstm, connolly2025does}. While these methods demonstrate the feasibility of forecasting some aspects of cross-platform dynamics, they are typically constrained to volume prediction or retrospective case studies~\cite{wilson2020cross, ng2022cross}. They rarely capture narrative-level emergence timing or provide platform-invariant representations suitable for real-time monitoring. As malicious actors continue to exploit fragmentation to adapt and seed narratives across ecosystems~\cite{minici2024uncovering, ng2022cross}, the need grows for predictive methods that are content-driven, robust to platform variation, and able to anticipate when and where narratives will emerge.




\section{Problem Formulation}

\subsection{Definitions}
\label{sec:definitions}

\vspace{2pt} \noindent
\textbf{Narrative.} Following prior work on narrative tracking~\cite{hanley2024specious,hanley2024partial, gerard2024modeling, hanley2023happenstance}, we define a \textit{narrative} as a collection of posts focusing on the same issue or event: a coherent unit of discourse that evolves across time and platforms. For instance, the Springfield narrative might include: ``Haitians eating pets in Ohio'' (Truth Social), ``immigrants consuming domestic animals - verified by locals!'' (Telegram), and ``Springfield residents confirm: pets are disappearing \#SpringfieldOhio'' (X). See Table~\ref{tab:narrative_examples}.


\vspace{2pt} \noindent
\textbf{Cross-Platform Emergence.} Following prior work~\cite{gerard2025bridging}, we define \textit{emergence} as occurring when a narrative appears on a source platform at least 48 hours before \textit{emerging} on a target platform with substantive adoption ($\geq$10 posts, the 35th percentile of narrative sizes)~\cite{nizzoli2021coordinated,magelinski2022synchronized}. The 48-hour threshold ensures temporal precedence rather than simultaneous reactions to external events. We operationalize emergence as a predictive signal: Platform B activity serves as an early indicator for forecasting emergence on Platform A, without claiming causal transmission between platforms.




    


\subsection{Prediction Task}

Narratives frequently appear on one platform before emerging on others: a pattern observed across misinformation~\cite{zannettou2019let}, coordinated campaigns~\cite{starbird2023influence}, and political discourse~\cite{ribeiro2021evolution}. Existing approaches generally fall into two categories: (1) diffusion simulations that model spreading dynamics but require extensive parameterization and platform-specific cascade data~\cite{pastor2015epidemic}, and (2) behavioral network methods that capture within-platform coordination but struggle to adequately bridge ecosystem boundaries~\cite{ng2022cross,luceri2024unmasking}. Both leave cross-platform prediction challenging.

We formalize cross-platform emergence as a prediction problem. Given a narrative $n$ active on source platform $s_{\text{src}}$ at time $t$, predict whether it will emerge on target platform $s_{\text{tgt}}$ within time window $\Delta$:

\begin{equation}
P\big(n \text{ emerges on } s_{\text{tgt}} \text{ by } t+\Delta \;\big|\; \text{activity on } s_{\text{src}} \text{ at } t\big)
\end{equation}

\noindent where emergence means reaching $\geq 10$ posts on the target platform. We test three horizons ($\Delta \in \{3, 7, 14\}$ days): 3-day predictions support reactive response, while 7-14 day forecasts enable proactive intervention.


\vspace{3pt}
\noindent

\section{Data}
\label{sec:data}

We construct a unified corpus of 5.73M posts from four datasets covering the 2024 U.S. presidential election: X~\cite{balasubramanian2024public}, TikTok~\cite{pinto2025tracking}, Truth Social~\cite{shah2024unfiltered}, and Telegram~\cite{blas2025unearthing}. The collection spans April–November 2024, comprising 991K distinct users (Table~\ref{tab:posts_by_platform}).

The data presents real-world heterogeneity across distinct platforms and modalities. TikTok comprises videos (analyzed via Whisper-generated\footnote{\url{https://github.com/openai/whisper}} transcripts)~\cite{pinto2025tracking}, while X, Truth Social, and Telegram provide native text~\cite{balasubramanian2024public,shah2024unfiltered,blas2025unearthing}. This heterogeneity---spanning affordances, user bases, and content modalities---captures the fragmented information landscape through which  modern narratives flow.



\subsection{Cross-Platform Content Extraction}

Cross-platform narrative tracking requires overcoming heterogeneity in content format and style. As shown in Table~\ref{tab:narrative_examples}, the same narrative appears in drastically different forms across platforms.

Following prior work~\cite{lawrence2020argument, li2025large, ku2024proceedings}, we address this through a large language model (LLM)-based claim extraction pipeline that distills heterogeneous content into semantically comparable units. Using Llama-3-70B-Instruct~\cite{dubey2024llama}, we extract claims and assertions from raw posts, removing platform-specific artifacts while preserving core semantic content. We validate this extraction in multiple, complementary ways. First, we evaluate semantic preservation using the UKP Sentential Argument Mining corpus~\cite{stab2018cross}, which contains expert-annotated argumentative claims across domains. Comparing our extracted claims to gold annotations using BERTScore~\cite{zhang2019bertscore}, the model achieves 0.82, confirming preserved semantic content. Second, we verify performance on our cross-platform data by having two annotators (substantial agreement, $\kappa=0.84$) evaluate whether 50 extracted claims per platform preserve original meaning. The system achieves 91.5\% accuracy across 200 claims. Finally, we test whether extractions are platform-invariant by training a BERT-based classifier to predict source platform from content. Platform discrimination accuracy drops from 0.67 (raw posts) to 0.26 (extracted claims), approaching random performance (0.25 for four platforms), confirming successful removal of platform-specific markers.


\subsection{Narrative Extraction}
Next, we follow established protocol for discovering narratives in text data: we embed the platform-invariant claims (using the Qwen3-Embedding~\cite{zhang2025qwen3} model, which we validate in Appendix~\ref{app:embedding_clustering} following protocols established in work~\cite{hanley2024specious, gerard2024modeling} )and cluster using DP-Means~\cite{hanley2024specious,hanley2024partial, gerard2024modeling, hanley2023happenstance}. DP-Means creates a new cluster whenever no existing center lies within $\lambda$ of a point, avoiding the need to pre-specify cluster counts. We use cosine distance threshold $\lambda = 0.10$ (similarity $\geq 0.90$), capturing issue-level narratives rather than broad themes. To validate clustering quality, we follow established protocols~\cite{hanley2024specious, gerard2024modeling}: two annotators ($\kappa=0.86$) evaluated whether randomly sampled posts within clusters discuss the same narrative. Across tested thresholds ($\lambda \in [0.075, 0.15]$), $\lambda=0.10$ achieves optimal performance (91.5\% accuracy). See Appendix~\ref{app:embedding_clustering}.

\subsection{Migratory Narratives}
\label{sec:migratory-narratives}
Building on the emergence definition introduced in Section~\ref{sec:definitions}, we identify which narrative clusters exhibit cross-platform emergence patterns. Of 10,679 clusters, 2,943 (27.6\%) appear on a source platform at least 48 hours before a target platform (with $\geq$10 posts on the target). While not causal, this temporal precedence establishes the prediction task: can we anticipate target platform emergence from source platform patterns?

To verify that temporal precedence reflects genuine predictive relationships rather than coincidental timing, we apply Granger causality testing. This tests whether source platform activity improves prediction of target platform emergence beyond what the target's own history explains. Of the 2,943 emergences, 2,056 (19.3\%) exhibit significant directional flow ($p < 0.05$, Bonferroni corrected, lag orders 2-7 days). Results are consistent across both definitions, so we report the more inclusive 2,943-migration set.



\begin{figure}[t] 
    \centering 
    \begin{overpic}[width=0.95\columnwidth]{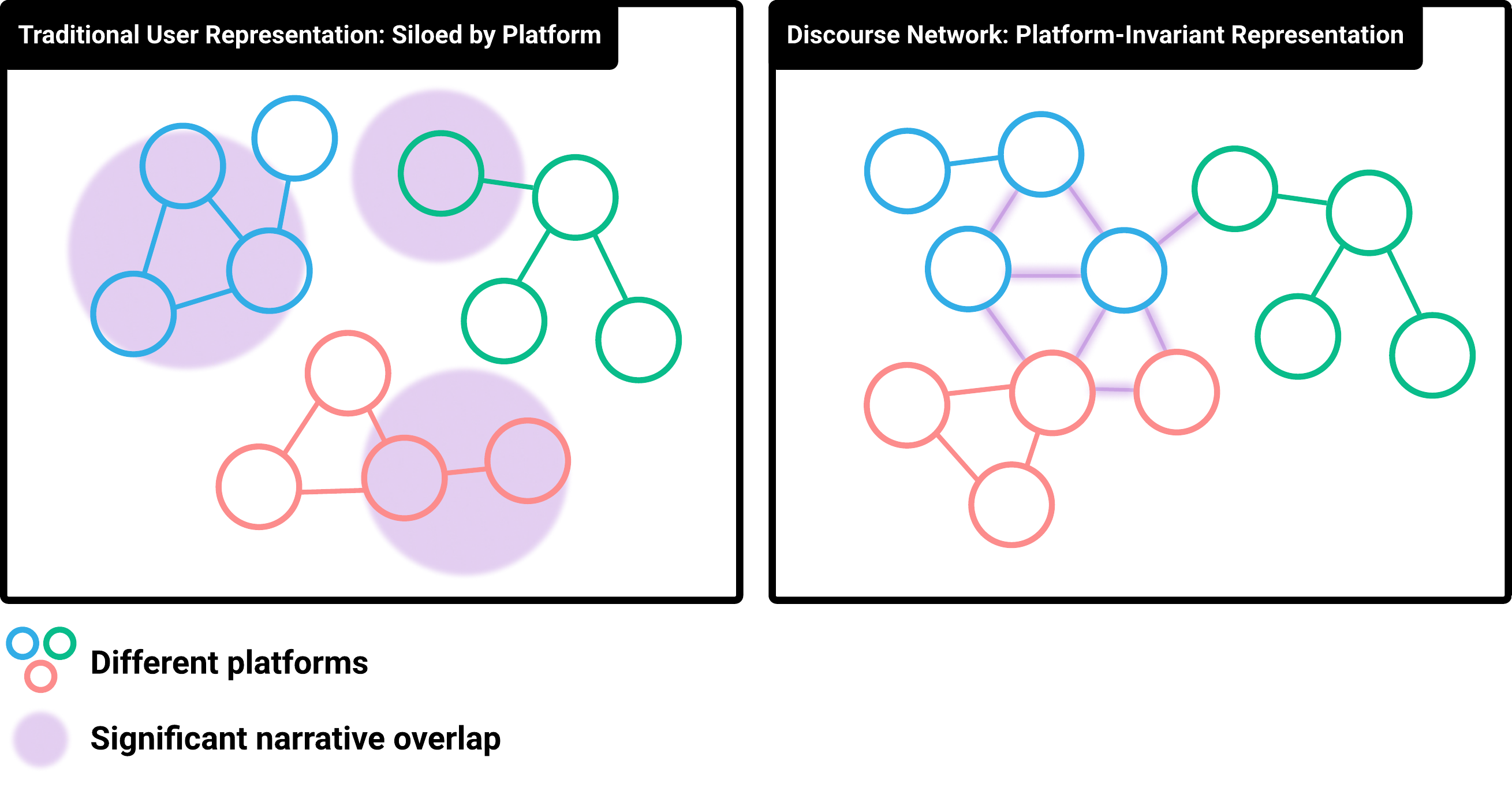}
    \end{overpic}
    \vspace{-1mm} 
\caption{Traditional user networks (left) fragment users by platform, while discourse networks (right) connect users through shared narrative engagement regardless of platform. Purple regions show cross-platform narrative overlap; node colors indicate platforms.}
    \vspace{-3mm}
    \label{fig:discourse-network}
\end{figure}

\section{Methodology}
\subsection{What are Discourse Networks?}
Consider two users discussing the Springfield narrative: one on Truth Social posted ``20,000 Haitians causing housing crisis'' in July, another on X posted ``Haitian migrants overwhelming infrastructure'' two days later. These users share no observable connection—no mutual follows, shared URLs, retweets, or platform overlap. Traditional network methods relying on follower links, repost cascades, or URL co-sharing cannot connect them~\cite{magelinski2022synchronized,pacheco2021uncovering,luceri2024unmasking}. Yet both engage with the same narrative at similar times, suggesting some form of latent alignment.

Discourse networks~\cite{gerard2025bridging} capture this alignment by representing users through shared narrative engagement rather than behavioral signals (Figure~\ref{fig:discourse-network}). If two users consistently engage with overlapping narratives---immigration concerns, infrastructure strain, election integrity---they become neighbors in the discourse network, regardless of platform. This platform-invariant representation connects users based on \textit{what they discuss} rather than \textit{how they interact}, mapping otherwise siloed platforms into a unified space where structural proximity reflects common discourse participation.

\subsection{Constructing Discourse Networks}

We construct discourse networks~\cite{gerard2025bridging} through three steps: \textit{narrative discovery} via clustering semantically similar posts, \textit{participation quantification} measuring each user's engagement with narratives, and \textit{network construction} connecting users with similar participation patterns.

\vspace{3pt}
\noindent
\textbf{Streaming Narrative Discovery.} We identify semantically coherent narratives using streaming hierarchical agglomerative clustering~\cite{gerard2024modeling}, which incrementally groups similar posts without pre-specifying cluster counts. Clusters refresh every 48 hours, and we use only clusters observed up to $t-2$ days to ensure temporal separation from prediction targets, mirroring deployment conditions where narrative counts are unknown and content arrives incrementally.

To avoid circularity with ground truth construction (which used DP-Means in Section~\ref{sec:data}), we employ a different clustering algorithm for prediction. Prior work shows discourse network performance depends on overall co-participation patterns rather than exact narrative boundaries~\cite{gerard2025bridging}, so this methodological separation likely does not compromise predictive power. For computational efficiency, all clustering uses MPNet-base-v2~\cite{song2020mpnet} embeddings (109M params vs. Qwen3's 596M params), providing substantial speedup while maintaining methodological separation from our dataset construction pipeline.





\vspace{3pt}
\noindent
\textbf{Claim Extraction Variants.}
To assess whether platform-specific language affects clustering, we test two discourse network variants. The first (\textit{status-based}) clusters raw post texts directly. The second (\textit{claim-based}) applies lightweight claim extraction using Gemma-4B-Instruct~\cite{team2025gemma} to normalize posts before embedding. We validate this model on the UKP Sentential Argument Mining corpus~\cite{stab2018cross}, achieving BERTScore 0.76 against gold annotations, confirming semantic preservation while removing stylistic variation.

We use Gemma-4B-Instruct rather than reusing Llama-3-70B-Instruct from our dataset construction for two reasons: (1) computational efficiency---4B parameters enables real-time processing versus 70B, and (2) methodological separation--using a different model family avoids potential circularity between dataset construction and prediction framework.

\vspace{3pt}
\noindent
\textbf{Network Construction.} Each user $u$ is represented by a vector of participation frequencies across narrative clusters $c$, weighted via TF-IDF transformation to emphasize distinctive engagement while down-weighting ubiquitous narratives. Users are then connected based on cosine similarity of their weighted participation vectors. Following prior work~\cite{gerard2025bridging}, we use TF-IDF weighting and cosine similarity as they best capture distinctive narrative engagement while remaining robust to high-volume users and imbalanced narratives. Full weighting formulation and justification appear in Appendix~\ref{sec:tf-idf}. Additionally, we analyze full computational requirements and deployment feasibility in Appendix~\ref{app:complexity_analysis}.

\subsection{Cross-Platform Social Proximity Principle}

Social proximity---the principle that structurally close individuals in networks exhibit correlated behaviors---is foundational to social network analysis~\cite{mcpherson2001birds, bollen2011happiness, barbera2015birds}. We hypothesize this principle extends to discourse networks as predictive correlation across platform boundaries. Unlike behavioral networks requiring direct interaction, discourse networks connect users through shared narrative engagement, potentially reflecting shared information exposure, ideological alignment, coordinated behavior, or other latent similarities—our data cannot distinguish between these mechanisms.

If discourse proximity captures some form of shared positioning in the information ecosystem, we should observe cross-platform correlation: users structurally close in the discourse network should exhibit correlated adoption patterns even on different platforms without observable connection. This correlation provides predictive signal without implying causal transmission.

Prior work demonstrates discourse networks perform well on cross-platform tasks such as engagement prediction and coordination detection~\cite{gerard2025bridging}, but the underlying social proximity mechanism has not been formally tested. We explicitly evaluate this through two hypotheses: \textbf{H1 (Within-platform):} Users show higher adoption rates when their same-platform discourse neighbors recently adopted a narrative. \textbf{H2 (Cross-platform):} Users show higher adoption rates when their cross-platform discourse neighbors recently adopted a narrative. These hypotheses motivate our predictive framework: we operationalize social proximity through neighbor activity features (Section~\ref{sec:operationalization}) and test whether the principle holds empirically in Section~\ref{sec:cross-platform-homophily}.

\subsection{Operationalizing Prediction}
\label{sec:operationalization}

The previous section posited that discourse networks should exhibit cross-platform social proximity: users are more likely to adopt a narrative when their discourse neighbors have recently done so. We now translate this principle into an operational framework for predicting cross-platform information diffusion, which we later validate empirically.

The intuition is straightforward: when discourse neighbors on Platform~A engage with a narrative, structurally proximate users on Platform~B are more likely to adopt it next. This reframes cross-platform prediction as a social proximity task rather than a diffusion simulation—prediction depends on monitoring activity among cross-platform neighbors.

Formally, for each target user $u$ on platform $s_{\text{tgt}}$ at time $t$, we identify their $k$ nearest cross-platform neighbors from source platform $s_{\text{src}}$ in the discourse network and compute features that capture recent neighbor activity. We extract two core measures: (1) \textbf{active connection count}, the number of cross-platform neighbors who engaged with the narrative within the past 7~days, and (2) \textbf{total connection count}, the total number of cross-platform neighbors. We also test the activity ratio (calculated as active connection count over  total connection count, though we find that the model implicitly captures this relationship during training. Temporal windows from 1–7~days yield consistent results, with a 7-day window providing the best balance between recency and coverage.

These features operationalize cross-platform social proximity as a simple neighbor-activity signal: prediction reduces to checking whether a user’s cross-platform neighbors have recently engaged with a narrative. Details on feature validation and network construction appear in Appendix sections~\ref{app:ablation} and~\ref{app:complexity_analysis}.



We train a Random Forest classifier using these features. Random Forest balances predictive performance with operational simplicity: requiring minimal tuning, providing interpretable feature importances, and demonstrating that proximity-based features alone suffice for cross-platform prediction without complex modeling. Notably, this approach scales efficiently: we need only compute features for active narratives and users, avoiding exhaustive pairwise comparisons. In Section~\ref{sec:ablation}, we further demonstrate the efficiency gains from this proximity-based reformulation: using network centrality to select users, we achieve equivalent performance while using only 2.9\% of active users (0.5\% of all observed users), substantially reducing computational and data collection requirements.


\section{Experiments}
\subsection{Baseline Approaches}
We compare discourse networks against three categories of approaches representing standard methods from the cross-platform diffusion literature~\cite{luceri2024unmasking,minici2024uncovering,ng2022cross}. 


\subsubsection{Diffusion Models}
We implement standard epidemic-style diffusion models adapted for cross-platform settings. These represent among the most established approaches to modeling information spread in networked systems~\cite{pastor2015epidemic, kempe2003maximizing}.

\vspace{3pt}
\noindent
\textbf{Hawkes processes}~\cite{hajiakhoond2019lstm} model narrative adoption as self-exciting point processes, where recent adoptions increase the probability of future adoptions. \textit{Implementation:} We fit separate Hawkes processes for each platform using narrative adoption timestamps, then use the learned excitation parameters to predict cross-platform jumps by treating source platform activity as exogenous input to target platform processes.

\vspace{3pt}
\noindent
\textbf{Independent cascade models} simulate information spread through network edges with fixed transmission probabilities. \textit{Implementation:} We adapt the model to cross-platform settings by constructing behavioral networks (when available) and treating cross-platform edges as lower-probability transmission paths ($p=0.1$ vs. $p=0.3$ within-platform), then run Monte Carlo simulations to estimate emergence probability.

\subsubsection{Content-Based Methods}

Non-network approaches represent standard temporal and popularity-based prediction methods that capture diffusion dynamics and have been directly tested in prior work~\cite{hajiakhoond2019lstm}.

\vspace{3pt}
\noindent
\textbf{Temporal sequence model (LSTM)} represent the most direct baseline for cross-platform prediction. ~\citet{hajiakhoond2019lstm} specifically developed LSTMs for predicting cross-platform bursts of social media activity. 
We adapt their platform-level burst prediction to narrative-level emergence by encoding activity as sequences of binary adoption events per user. Models use hidden dimension 128 and 3 layers, with hyperparameters tuned via grid search on validation data (full specifications in repository).

\vspace{3pt}
\noindent
\textbf{Popularity-based prediction} assumes narratives gaining momentum on source platforms are more likely to spread, capturing viral diffusion patterns independent of user connections. \textit{Implementation:} For each narrative on the source platform, we compute growth rate (posts per hour), engagement velocity (likes/shares per post over time), and adoption curve steepness. These features predict whether the narrative will reach the target platform within each time horizon.

\vspace{3pt}
\noindent
\textbf{Platform transition matrices} assume platform-level flow probabilities can predict individual emergence events without user-level factors. \textit{Implementation:} We compute historical emergence frequencies between all platform pairs (e.g., Truth Social $\rightarrow$ X, Telegram $\rightarrow$ TikTok) and use these transition probabilities to predict whether a narrative active on platform A will appear on platform B within each time window.

\subsubsection{Behavioral Network-Based Methods}

For fair comparison, we implement standard network construction methods from cross-platform analysis literature~\cite{magelinski2022synchronized,pacheco2021uncovering,luceri2024unmasking}. We implement: \textbf{Co-URL networks} connect users who share the same URLs in their posts, identifying communities coordinating around shared information sources. \textbf{Hashtag sequence networks} connect users based on ordered sequences of shared hashtags, capturing strategic tagging behavior often associated with coordinated campaigns~\cite{burghardt2024socio}. \textbf{Text similarity networks} connect users if they post at least one highly similar post (cosine similarity $> 0.8$), with edge weights reflecting average text similarity across matches~\cite{pacheco2021uncovering}. \textbf{k-NN embedding networks} build a post-to-post k-nearest neighbor graph based on embedding similarity, then induce user-to-user connections reflecting overall proximity across all posts~\cite{ng2023coordinating}. \textbf{Fused networks} construct a unified network where users are linked if they are connected in any underlying similarity network (Co-URL, Hashtag Sequence, or Text Similarity)~\cite{luceri2024unmasking}.


\subsection{Evaluation Protocol}



\vspace{3pt}
\noindent
\textbf{Emergence Detection Ground Truth.} Using migratory narratives from Section~\ref{sec:migratory-narratives}, we construct binary labels for prediction. For each narrative $n$ active on source platform $s_{\text{src}}$ at time $t$, we create labels for target platform $s_{\text{tgt}}$ and horizons $\Delta \in \{3,7,14\}$ days. A positive label indicates the narrative reached adoption threshold ($\geq$10 posts) on $s_{\text{tgt}}$ within $[t, t+\Delta]$. Note that ``source'' denotes first observation in our data rather than definitive narrative origin. Table~\ref{tab:task1_distribution} shows class distributions across horizons, reflecting real-world imbalance where many narratives remain confined to their initial platform short-term.

\vspace{3pt}
\noindent
\textbf{Training Details.} 

\noindent
\textit{Discourse Network Model.} We train a Random Forest classifier on  proximity features (Section~\ref{sec:operationalization}) using 100 estimators with balanced class weights. Features are z-score normalized using mean and standard deviation computed from the training set only, then applied to validation and test sets.

\noindent
\textit{Baseline Models.} Network-based baselines extract the same proximity features and train Random Forest classifiers with identical hyperparameters for fair comparison. For semantic baselines (Text Similarity, k-NN Embedding), we test both raw posts and claim-extracted text (using claims extracted by the same Gemma-4B-Instruct model used for discourse network construction), reporting whichever performs better to isolate network construction effects from normalization benefits. Temporal sequence model (LSTM) uses hidden dimension 128 and 3 layers, with hyperparameters tuned via grid search on validation data. Diffusion models (Hawkes, IC) use fixed parameters as they approximate theoretical dynamics rather than fit to prediction tasks. See repository for full implementation.

\vspace{3pt}
\noindent
\textbf{Evaluation Protocol.} We implement streaming evaluation replicating real-world deployment. At each prediction time $t$ in our test period (October-November 2024), we train models using all data from April through $t$, then predict narrative emergence for $t + \Delta$ days. Networks and features use only information available at $t$. When predicting emergence from $s_{\text{src}}$ to $s_{\text{tgt}}$, we exclude data from $t$ onward on $s_{\text{tgt}}$ to prevent information leakage. We train separate models for each horizon (3d, 7d, 14d).

\vspace{3pt}
\noindent
\textbf{Evaluation Metrics.} We To evaluate predictive accuracy, we report \textbf{AUC}, \textbf{F1}, and \textbf{Precision} (fraction of positive predictions that are correct, crucial for operational systems where false alarms waste analyst resources). We omit Recall from Table~\ref{tab:migration_detection} for clarity, as it provides the least additional information given the other metrics and can be derived from F1 and Precision. Full results including Recall are available in the repository.

To evaluate practical utility, we complement these standard metrics with \textit{operator-oriented} measures. \textbf{Precision@k} assesses how accurately the model identifies the most actionable subset of emergence events; this mirrors real-world analyst triage under limited review budgets. \textbf{Cumulative gain} (yield curve) measures the proportion of true cross-platform emergences captured as analysts review increasing fractions of the ranked precisions; this reflects operational tradeoffs between coverage and workload.

\begin{table*}[t]
\centering
\caption{Multi-horizon emergence detection performance. Reported are AUC, F1, and Precision scores (mean $\pm$ std) across prediction windows.}
\label{tab:migration_detection}
\center
\resizebox{1.6\columnwidth}{!}{%
\begin{tabular}{l|ccc|ccc|ccc}
\toprule
 & \multicolumn{3}{c|}{\textbf{AUC}} & \multicolumn{3}{c|}{\textbf{F1}} & \multicolumn{3}{c}{\textbf{Precision}} \\
\textbf{Method} & \textbf{3d} & \textbf{7d} & \textbf{14d} & \textbf{3d} & \textbf{7d} & \textbf{14d} & \textbf{3d} & \textbf{7d} & \textbf{14d} \\
\midrule
\multicolumn{10}{l}{\textit{No Network Baselines}} \\
Popularity Baseline & $0.50_{\pm 0.00}$ & $0.50_{\pm 0.00}$ & $0.50_{\pm 0.00}$ & $0.24_{\pm 0.00}$ & $0.28_{\pm 0.00}$ & $0.31_{\pm 0.00}$ & $0.13_{\pm 0.00}$ & $0.16_{\pm 0.00}$ & $0.18_{\pm 0.00}$ \\
Platform Transitions & $0.62_{\pm 0.00}$ & $0.79_{\pm 0.00}$ & $0.76_{\pm 0.00}$ & $0.45_{\pm 0.00}$ & $0.53_{\pm 0.00}$ & $0.54_{\pm 0.00}$ & $0.36_{\pm 0.00}$ & $0.43_{\pm 0.00}$ & $0.43_{\pm 0.00}$ \\
LSTM (engagement) & $0.50_{\pm 0.00}$ & $0.50_{\pm 0.00}$ & $0.50_{\pm 0.00}$ & $0.24_{\pm 0.02}$ & $0.29_{\pm 0.00}$ & $0.31_{\pm 0.01}$ & $0.14_{\pm 0.00}$ & $0.17_{\pm 0.01}$ & $0.19_{\pm 0.01}$ \\
\cmidrule(lr){1-10}
\multicolumn{10}{l}{\textit{Diffusion Models}} \\
Hawkes Process & $0.57_{\pm 0.00}$ & $0.56_{\pm 0.01}$ & $0.55_{\pm 0.001}$ & $0.26_{\pm 0.01}$ & $0.28_{\pm 0.00}$ & $0.32_{\pm 0.00}$ & $0.18_{\pm 0.00}$ & $0.21_{\pm 0.00}$ & $0.19_{\pm 0.00}$ \\
Independent Cascade & $0.58_{\pm 0.00}$ & $0.57_{\pm 0.00}$ & $0.56_{\pm 0.00}$ & $0.26_{\pm 0.00}$ & $0.28_{\pm 0.00}$ & $0.32_{\pm 0.0}$ & $0.17_{\pm 0.00}$ & $0.19_{\pm 0.00}$ & $0.19_{\pm 0.00}$ \\
\cmidrule(lr){1-10}
\multicolumn{10}{l}{\textit{Other Networks}} \\
Co-URL & $0.50_{\pm 0.01}$ & $0.50_{\pm 0.00}$ & $0.50_{\pm 0.00}$ & $0.25_{\pm 0.01}$ & $0.24_{\pm 0.01}$ & $0.23_{\pm 0.01}$ & $0.15_{\pm 0.03}$ & $0.13_{\pm 0.00}$ & $0.13_{\pm 0.01}$ \\
Hashtag Sequence & $0.51_{\pm 0.03}$ & $0.50_{\pm 0.00}$ & $0.50_{\pm 0.00}$ & $0.24_{\pm 0.03}$ & $0.26_{\pm 0.01}$ & $0.23_{\pm 0.00}$ & $0.15_{\pm 0.02}$ & $0.14_{\pm 0.00}$ & $0.13_{\pm 0.01}$ \\
Text Similarity & $0.61_{\pm 0.02}$ & $0.72_{\pm 0.01}$ & $0.76_{\pm 0.01}$ & $0.40_{\pm 0.01}$ & $0.50_{\pm 0.01}$ & $0.47_{\pm 0.00}$ & $0.32_{\pm 0.01}$ & $0.36_{\pm 0.00}$ & $0.35_{\pm 0.00}$ \\
k-NN Embedding & $0.65_{\pm 0.01}$ & $0.71_{\pm 0.00}$ & $0.76_{\pm 0.01}$ & $0.41_{\pm 0.02}$ & $0.52_{\pm 0.02}$ & $0.47_{\pm 0.01}$ & $0.32_{\pm 0.02}$ & $0.37_{\pm 0.01}$ & $0.36_{\pm 0.01}$ \\
Fused Network & $0.69_{\pm 0.01}$ & $0.74_{\pm 0.01}$ & $0.79_{\pm 0.02}$ & $0.44_{\pm 0.04}$ & $0.52_{\pm 0.01}$ & $0.49_{\pm 0.01}$ & $0.34_{\pm 0.01}$ & $0.39_{\pm 0.00}$ & $0.35_{\pm 0.02}$ \\
\cmidrule(lr){1-10}
\multicolumn{10}{l}{\textit{Discourse Network}} \\
No Claim Extraction & $0.79_{\pm 0.01}$ & $0.80_{\pm 0.01}$ & ${0.80_{\pm 0.00}}$ & ${0.49_{\pm 0.01}}$ & ${0.58_{\pm 0.01}}$ & ${0.59_{\pm 0.02}}$ & ${0.49_{\pm 0.00}}$ & ${0.45_{\pm 0.01}}$ & ${0.48_{\pm 0.02}}$ \\
Claim Extraction & $\underline{0.88_{\pm 0.01}}$ & $\underline{0.87_{\pm 0.01}}$ & $\underline{0.86_{\pm 0.02}}$ & $\underline{0.58_{\pm 0.02}}$ & $\underline{0.61_{\pm 0.02}}$ & $\underline{0.62_{\pm 0.01}}$ & $\underline{0.56_{\pm 0.02}}$ & $\underline{0.48_{\pm 0.01}}$ & $\underline{0.54_{\pm 0.01}}$ \\
+ Platform Transitions & $\bm{0.94_{\pm 0.02}}$ & $\bm{0.94_{\pm 0.01}}$ & $\bm{0.92_{\pm 0.01}}$ & $\bm{0.66_{\pm 0.01}}$ & $\bm{0.72_{\pm 0.03}}$ & $\bm{0.68_{\pm 0.02}}$ & $\bm{0.65_{\pm 0.03}}$ & $\bm{0.63_{\pm 0.01}}$ & $\bm{0.74_{\pm 0.01}}$ \\
\cmidrule(lr){1-10}
Random Baseline & $0.49_{\pm 0.03}$ & $0.50_{\pm 0.02}$ & $0.53_{\pm 0.02}$ & $0.24_{\pm 0.03}$ & $0.28_{\pm 0.01}$ & $0.31_{\pm 0.03}$ & $0.14_{\pm 0.03}$ & $0.24_{\pm 0.04}$ & $0.19_{\pm 0.03}$ \\
\bottomrule
\end{tabular}
}
\end{table*}

\begin{table}[h]
\centering
\caption{
Precision@1\% and Precision@5\% for the best-performing model in each family on the 7-day prediction window.
Values are stable across horizons (3 d, 7 d, 14 d); complete results are available in the project repository.
}
\label{tab:precision_top_models}
\begin{tabular}{lcc}
\toprule
\textbf{Model Family} & \textbf{P@1\%} & \textbf{P@5\%} \\
\midrule
Random (Class Baseline) & $0.16$ & $0.16$ \\
No Network (Platform Transitions) & $0.32$ & $\underline{0.45}$ \\
Diffusion (Independent Cascade) & $\underline{0.35}$ & $0.26$ \\
Traditional Networks (Fused) & $0.28$ & $0.34$ \\
\textbf{Discourse Network (Ours)} & $\bm{0.80}$ & $\bm{0.60}$ \\
\bottomrule
\end{tabular}
\end{table}

\subsection{Validating Cross-Platform Social Proximity}
\label{sec:cross-platform-homophily}

We test whether users who are socially proximate---that is, whose discourse neighbors recently engaged with a narrative---are more likely to adopt that narrative themselves. This analysis evaluates whether network-based social proximity provides predictive signal for narrative adoption both within and across platforms. We measure, for each user, the proportion of active discourse neighbors ($R$) who have already engaged with the narrative, and estimate how adoption likelihood varies with $R$. Details on model construction, regression specification, and robustness checks are provided in Appendix~\ref{app:validation-suite-social-proximity}.

Adoption probability increases monotonically with $R$. Users in the bottom quintile of $R$ adopt with probability $0.14$, compared to $0.21$ in the top quintile---a 52.7\% lift (95\% CI: [40.6, 61.4]). Finer-grained bins confirm the same pattern: adoption rates are $13.5\%$ with no active neighbors, $17.4\%$ with one to three, and $21.7\%$ with six or more. To account for user activity, platform, and time effects, we estimate a logistic regression predicting adoption as a function of $R$. Users with active discourse neighbors have odds of adoption $22.4\times$ higher (95\% CI: [16.9, 29.7]) than those without, even after adjusting for these controls. Both within-platform and cross-platform neighbor activity remain significant predictors, demonstrating that social proximity captures meaningful behavioral alignment across platform boundaries. For comparison, we replicate the analysis using the fused network setup. While still significant, its effect is markedly weaker (\textit{OR} = 1.57, 95\% CI = [1.54–1.60]), indicating that explicitly modeling cross-platform discourse edges captures stronger and more localized proximity effects.

These results support \textit{cross-platform social proximity}: users connected through discourse structures adopt narratives in correlated fashion, even without direct interaction.

\subsection{Emergence Detection Performance}
\label{sec:results-multihorizon}

\vspace{3pt}
\noindent
\textbf{Overall Performance.} Table~\ref{tab:migration_detection} shows discourse networks substantially outperform all baselines across prediction horizons: 27\% AUC improvement and 32\% F1 improvement over the next-best baseline, and 54\% AUC improvement over diffusion models. Performance remains stable across 3-, 7-, and 14-day windows, indicating discourse signals support multi-day forecasting. These gains likely stem from reformulating diffusion as social proximity: when emergence occurs, users have 290\% more active cross-platform neighbors than when emergence does not occur. This operationalizes a simple question: are structurally similar users on other platforms already discussing this topic? Adding historical platform-pair emergence patterns (e.g., P(TikTok $\rightarrow$ Twitter) from past data) provides modest additional gains by contextualizing which platform pairs exhibit frequent emergence.

The fused network represents the strongest traditional approach. However, 99.4\% of its edges originate from text similarity, reflecting behavioral methods' reliance on near-exact matches. Figure~\ref{fig:ccdf-cap} illustrates the consequence: fused networks concentrate ties among hubs (median cross-degree 1, mean 1.9, 50\% disconnected), while discourse networks exhibit widespread connectivity (median 9, mean 14.9, 93\% connected). This allows discourse-based prediction to leverage collective behavior of ordinary users rather than central hubs.

\vspace{3pt}
\noindent
\textbf{Operational Utility.} Table~\ref{tab:precision_top_models} reports Precision@k on highest-ranked predictions. Discourse networks achieve 80\% precision on the top 1\% and 60\% on the top 5\%, concentrating true emergences where analysts look first. Figure~\ref{fig:cum_gain_rf} shows cumulative gain: discourse networks recover 40\% of emergences by reviewing just 5\% of predictions, reaching 85\% recall at 30\% budget. The top 5\% of predictions correctly identify 60\% of flagged cases while capturing 40\% of all emergences: actionable early warning with manageable workload.


\subsection{Ablation Studies and Feature Analysis}
\label{sec:ablation}
\vspace{3pt}
\noindent
\textbf{Feature Importance.} Table~\ref{tab:ablation} evaluates component contributions to prediction performance. Among individual features, active connection count (number of cross-platform neighbors who recently engaged with the narrative at hand) achieves highest performance ($AUC=0.77$). This combined with total cross-platform connection count achieves peak performance ($AUC=0.88$), indicating that the model benefits from both from knowing the number of active neighbors as well as the  total number possible. This combined approach outperforms activity ratio alone (calculating the proportion of active neighbors, $AUC=0.73$), indicating that absolute counts provide richer signal than pre-computed proportions, and suggesting the model implicitly learns proportional relationships during training. Claim extraction provides substantial gains, improving what would be maximum performance of $AUC=0.79$ to $0.88$. Finally, historical patterns grant additional gains ($+0.06$ AUC), as they likely provide directional context for narrative movement.

\definecolor{Diff}{HTML}{EF5B5B}

\begin{figure}[htbp]
\begin{tikzpicture}
\hspace{-10pt}
\begin{axis}[
  width=0.98\linewidth, height=0.68\linewidth,
  xlabel={Annotation budget (\%)}, ylabel={Recall},
  grid=both, grid style={opacity=0.1},
  xmin=0, xmax=30, ymin=0, ymax=1,
  legend style={
    at={(0.02,0.98)},
    anchor=north west,
    draw=none,
    fill=none,
    font=\tiny,
    align=left
  },
  legend cell align=left
]
\addplot[Bridge, thick, mark=*, mark size=.2pt]
  table[x=pct_reviewed, y=recall, col sep=comma]{csvs/cumulative_gain_rf_thin.csv};
\addlegendentry{Discourse (Ours)}

\addplot[NoBridge, thick, mark=*, mark size=.2pt]
  table[x=pct_reviewed, y=recall, col sep=comma]{csvs/cumulative_gain_nn_thin_fused.csv};
\addlegendentry{Second Best (Fused Network)}

\addplot[Diff, thick, mark=*, mark size=.2pt]
  table[x=pct_reviewed, y=recall, col sep=comma]{csvs/cumulative_gain_hawkes_thin.csv};
\addlegendentry{Best Diffusion (Hawkes)}

\addplot[black, thick, dotted, domain=0:30, samples=2] ({x},{x/100});
\addlegendentry{Random Baseline}

\end{axis}
\end{tikzpicture}
\caption{Cumulative gain (operator yield curve) comparing Discourse and Fused networks on the 7-day prediction window. The dashed line indicates random performance.}
\label{fig:cum_gain_rf}
\end{figure}
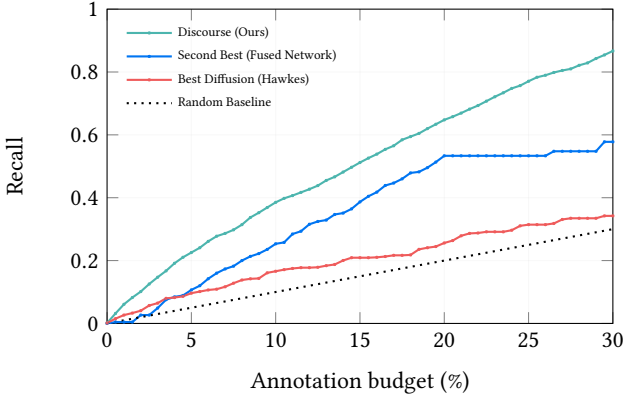

\vspace{3pt}
\noindent\textbf{Which Users Give the Most Signal?} We compare three user selection strategies: \textit{random downsampling} (baseline testing uniform signal distribution~\cite{leskovec2006sampling}), \textit{popularity-based selection} (highly visible accounts following influence maximization principles~\cite{kempe2003maximizing}), and \textit{cross-platform connectivity} (users ranked by inter-platform degree: number of connections to users on other platforms~\cite{freeman1978centrality}, hypothesized to function as bridge users across platform boundaries~\cite{granovetter1983strength,gerard2025bridging}). 

Cross-platform connectivity proves substantially more efficient: per user removed, it causes 8$\times$ less performance degradation than random sampling and 4$\times$ less than popularity. Notably, these bridge users exhibit median engagement levels (post count: 52nd percentile, likes: 53rd, replies: 51st)—their signal stems from network position, not visibility. This aligns with prior findings that users ``bridging'' platform discourse provide disproportionate predictive signal~\cite{gerard2025bridging}.

Operationally, using only the top quartile of cross-platform users (2.9\% of active users) retains 97\% of baseline performance, demonstrating scalable real-time detection across platforms.

\section{Retrospective Deployment Study}

To demonstrate operational feasibility, we deploy the full system as a self-contained pipeline processing all 5.7M posts chronologically from April-November 2024. The system operates with minimal configuration: clustering threshold $\lambda=0.10$, 48-hour update cycles, and the emergence definition established in Section~\ref{sec:definitions}. All other components: claim extraction normalization, clustering, network construction, prediction, run automatically.

\vspace{2pt} \noindent
\textbf{System Implementation.} The pipeline processes posts in four stages: (1) Gemma-4B-Instruct claim extraction at 245 posts/second, (2) MPNet-base-v2 embedding generation, (3) incremental clustering~\cite{gerard2024modeling}, and (4) discourse network construction. The system identifies past migratory narratives from historical data using the definition from Section~\ref{sec:definitions}, then trains a Random Forest model on proximity features from the top quartile of structurally central users using these narratives as pseudo-labels, and predicts narrative emergence across horizons $\Delta \in \{3,7,14\}$ days. Full architecture details and computational requirements appear in Appendix~\ref{app:complexity_analysis}.

\vspace{2pt} \noindent
\textbf{Operational Performance.} The system achieves stable performance within 24 days, maintaining AUC and F1 scores consistent with controlled experiments (Section~\ref{sec:results-multihorizon}). Rather than re-reporting quantitative metrics, we present case studies demonstrating early detection capabilities: which narratives received high-risk scores, how many days before mainstream emergence, and what intervention windows existed. We present two detailed examples below; Table~\ref{tab:notable_narratives_expanded} in the appendix contains additional high-impact narratives the system correctly predicted.

\vspace{2pt} \noindent
\textbf{Case Study: Springfield Conspiracy.} The ``Haitians eating pets'' conspiracy (July-September 2024) illustrates early detection capabilities. On September 2, 2024, our system identified this narrative on Telegram and correctly forecasted X/Twitter emergence within 3 days. The prediction proved accurate: according to external reporting beyond our dataset~\footnote{See \href{https://www.nbcnews.com/politics/donald-trump/trump-fringe-online-claim-immigrants-eating-pets-debate-trump-rcna170759}{NBC News report}}, the claim began circulating on X via Facebook screenshots on September 5, reached 1,100 posts by September 6, and surged to 9,000+ posts by September 7 (+720\%). The narrative was subsequently amplified by prominent political figures, including references during the September 10 presidential debate viewed by over 67M people.


\vspace{2pt} \noindent
\textbf{Case Study: Hurricane Helene Conspiracy.} The FEMA obstruction conspiracy (late September-October 2024) demonstrates detection of crisis-pivoted narratives. Our system identified claims on Telegram that federal responders were blocking aid and seizing supplies, correctly forecasting X emergence within 7 days. The prediction proved accurate: according to external reporting~\footnote{See \href{https://www.theguardian.com/us-news/2024/oct/14/north-carolina-hurricane-helene-fema-armed-militia-threat}{Guardian report}}, the conspiracy spread rapidly across platforms, reaching substantial volume on X by early October.

\section{Discussion}

\vspace{3pt}
\noindent
\vspace{3pt}
\noindent
\textbf{Cross-Platform Proximity as a Unifying Framework.} Discourse networks exhibit predictive social proximity across platforms, substantially outperforming platform-specific behavioral signals and diffusion models. This resolves a core challenge: narratives spread between platforms, yet existing methods struggle to capture this movement. Diffusion models treat cross-platform spread as exogenous shocks, while behavioral signals (URLs, hashtags, repost networks) break across platform boundaries. 
Discourse networks unify fragmented ecosystems by connecting users through narrative co-engagement regardless of platform, allowing standard proximity principles to predict emergence: narratives appear where active connections exist. This makes cross-platform prediction tractable without platform-specific engineering—the same network construction applies across text (X), video (TikTok), forwarded messages (Telegram), and mixed media (Truth Social).




\vspace{3pt}
\noindent
\textbf{Practical Implications for Detection Systems.} Cross-platform detection systems can achieve both efficiency and accuracy by focusing on structurally important users. Using proximity features from only 2.9\% of active accounts---those with high cross-platform connectivity---substantially outperforms methods requiring comprehensive behavioral data. At operational thresholds, discourse networks achieve substantially higher precision than baselines, concentrating true emergences where analysts investigate first. Cumulative gain curves demonstrate efficient resource allocation: discourse networks recover 40\% of emergences by reviewing just 5\% of predictions, enabling actionable early warning with manageable analyst workload while reducing data requirements.

\vspace{3pt}
\noindent
\textbf{Ethical Statement.} All data comes from publicly available datasets with institutional review~\cite{balasubramanian2024public, pinto2025tracking, shah2024unfiltered, blas2025unearthing}. For added security, we anonymize user identifiers via random hashing and release code without user-level identifiers. This work supports fact-checking organizations and platform safety teams in detecting emergent narratives and rumors for verification. We acknowledge dual-use risks, however: the same methods that we intend to help build healthier, safer discourse could support surveillance of individuals or communities. Finally, our system predicts emergence patterns without judging the content itself: determining appropriate interventions requires human expertise and contextual judgment beyond our technical framework.





\vspace{3pt}
\noindent
\textbf{Limitations.} Our coverage is incomplete: we analyze X, TikTok, Truth Social, and Telegram but omit platforms such as Reddit, YouTube, Discord, and international sites. Broader coverage would test whether discourse network principles hold across platforms with different affordances (e.g., threaded forums, long-form video, ephemeral chat). Our scope is also temporal. Data from the 2024 U.S. election may not reflect typical ecosystems, since elections involve unusual coordination and emergence. Testing other contexts would help assess generalizability. Finally, we predict correlation, not causation. High active connection ratios show where narratives appear, but not why: whether due to shared traits, coordination, or common media diets. Causal inference would help disentangle mechanisms and guide interventions.

\vspace{3pt}
\noindent
\textbf{Future Work.} Future work should address real-time deployment challenges. Our batch predictions assume narrative clusters already exist; operational systems must detect emerging narratives continuously while maintaining low false positive rates. Broader platform coverage, longer temporal validation, and mechanistic understanding would strengthen the approach's practical utility for monitoring fragmented information ecosystems.

\section{Conclusion}
Online narratives diffuse unevenly across platforms, yet existing methods treat each platform as an isolated system, relying on platform-specific signals that cannot capture cross-platform emergence. We address this by constructing platform-invariant discourse networks that link users through shared narrative engagement, creating a unified representation where cross-platform proximity predicts narrative emergence.

On 5.7M posts from X, TikTok, Truth Social, and Telegram during the 2024 U.S. election, proximity-based features substantially outperform diffusion simulations and behavioral network baselines (AUC 0.88 for 3-14 day prediction). Using only 2.9\% of active users retains 97\% of performance, demonstrating efficiency. Applied retrospectively, this approach anticipated the Hurricane Helene FEMA conspiracy and Springfield rumors days before mainstream emergence.

By representing users through narrative co-engagement rather than platform-specific behaviors, social proximity principles can be extended across platform boundaries. This reformulation transforms cross-platform diffusion from a complex multi-platform coordination problem into tractable neighbor activity monitoring, achieving strong predictive performance without requiring comprehensive behavioral data, platform-specific engineering, or complex diffusion models.

\clearpage
\bibliography{main}
\clearpage
\appendix
\section{Embedding Model Selection and Clustering Validation}
\label{app:embedding_clustering}

To select an appropriate embedding model for narrative clustering, we compare two candidates: MPNet-base-v2~\cite{song2020mpnet} and Qwen3-Embedding~\cite{zhang2025qwen3}. We evaluate each model's ability to capture narrative coherence through human annotation (following prior works' protocols~\cite{hanley2024specious, gerard2025bridging}) at multiple clustering thresholds. For each threshold $\lambda$ (e.g., 0.10), we: (1) cluster extracted claims using DP-Means with that threshold; (2) sample 50 random post pairs from the same cluster (each pair is from the same cluster, we sample from 50 clusters); (3) have human annotators judge whether each pair discusses the same narrative; (4) compute agreement as the proportion of correct clustering decisio.ns

Table~\ref{tab:embedding_comparison} shows human agreement across thresholds for both embedding models.

\begin{table}[h]
\centering
\caption{Human evaluation accuracy for narrative clustering across embedding models and thresholds.}
\small
\begin{tabular}{lll}
\toprule
\textbf{Threshold ($\lambda$)} & \textbf{MPNet-base-v2} & \textbf{Qwen3-Embedding} \\
\midrule
0.075 & 92\% & 94.5\% \\
0.10 & 88\% & 91.5\% \\
0.125 & 84\% & 88\% \\
0.15 & 80\% & 84\% \\
\bottomrule
\end{tabular}
\label{tab:embedding_comparison}
\end{table}

We select Qwen3-Embedding based on its superior performance (91.5\% vs. 88\% at $\lambda=0.10$) and greater stability across thresholds. The $\lambda=0.10$ threshold balances precision (avoiding false merges of distinct narratives) with recall (capturing narratively equivalent posts despite linguistic variation). Stricter thresholds (e.g., $\lambda=0.075$) achieve higher accuracy but risk over-segmentation, while looser thresholds ($\lambda=0.15$) merge distinct narratives. 





\section{TF-IDF Weighting for Discourse Networks}
\label{sec:tf-idf}

We represent each user $u$ as a vector of participation frequencies across narrative clusters $c$. To weight these associations, we apply TF-IDF (Term Frequency-Inverse Document Frequency) transformation, which emphasizes distinctive engagement while down-weighting ubiquitous narratives:

\[
w_{u,c} = \text{tf}_{u,c} \cdot \log\!\left(\frac{|U|}{|\{u : u \in c\}|}\right)
\]

where:
\begin{itemize}
    \item $\text{tf}_{u,c}$: number of posts by user $u$ in cluster $c$ (term frequency)
    \item $|U|$: total number of users in the network
    \item $|\{u : u \in c\}|$: number of users who engaged with narrative $c$ (document frequency)
\end{itemize}

\vspace{3pt}
\noindent
\textbf{Rationale.} TF-IDF weighting serves two purposes. First, it emphasizes distinctive narratives: users heavily engaged with rare narratives receive higher weights than those discussing ubiquitous topics. Second, it normalizes for narrative popularity: a user posting 10 times about a niche narrative (engaged by 100 users) receives higher weight than posting 10 times about a mainstream narrative (engaged by 10,000 users).

Users are then connected via cosine similarity of their weighted participation vectors:

\[
\text{sim}(u_i, u_j) = \frac{\mathbf{w}_i \cdot \mathbf{w}_j}{\|\mathbf{w}_i\| \|\mathbf{w}_j\|}
\]

where $\mathbf{w}_i$ is user $i$'s TF-IDF weighted participation vector. Cosine similarity provides scale-invariant comparison, ensuring users with different activity levels can be compared: a user posting 10 times with the same narrative distribution as someone posting 1,000 times will have high similarity.

\vspace{3pt}
\noindent
\textbf{Alternative weighting schemes.} We follow Gerard et al.~\cite{gerard2025bridging}, who compared TF-IDF against the following alternatives: \textit{Raw frequency}: Uses unnormalized counts of user participation in each narrative cluster; \textit{Softmax normalization}: Applies softmax transformation over per-user cluster participation counts. Their empirical evaluation found TF-IDF + cosine similarity optimal for capturing meaningful user alignment in discourse networks while remaining robust to activity level variation and narrative imbalance.

To simulate real-world conditions, we use use FAISS-HNSW approximate nearest neighbor search~\cite{malkov2018efficient}, reducing complexity from $O(n^2 d)$ to $O(n \log n \cdot d)$. This has been validated in prior work~\cite{gerard2025bridging}. 

\section{Logistic Regression Validation}
\label{app:validation-suite-social-proximity}

To evaluate whether social proximity effects persist after accounting for user- and platform-level heterogeneity, we estimate a logistic regression predicting narrative adoption from the proportion of active discourse neighbors ($R$). The model specification is:

\begin{equation}
\text{logit}(P(\text{adopt}_{i,n,t})) = \beta_0 + \beta_1 R_{i,n,t} + \beta_2 \log(1 + \text{posts}_i) + \gamma_p + \delta_t
\end{equation}

where $i$ indexes users, $n$ indexes narratives, $t$ indexes time periods, $R_{i,n,t}$ is the proportion of user $i$'s discourse neighbors who have engaged with narrative $n$ by time $t$, $\log(1 + \text{posts}_i)$ controls for user activity level, $\gamma_p$ are platform fixed effects, and $\delta_t$ are time fixed effects. We fit the model using maximum likelihood with robust standard errors clustered at the user level.

\vspace{3pt}
\noindent
\textbf{Platform-specific effects.} Because platforms differ in structure and behavior norms, we estimate platform-specific coefficients via interaction terms: $\beta_1 \cdot R \cdot \mathbb{1}[\text{platform} = p]$. To compute an overall effect size, we take a weighted average of platform-specific odds ratios, weighted by each platform's prevalence in the sample:

\begin{equation}
\text{OR}_{\text{weighted}} = \exp\left(\sum_{p} w_p \cdot \beta_{1,p}\right)
\end{equation}

where $w_p$ is the proportion of observations on platform $p$, and $\beta_{1,p}$ is the platform-specific log-odds coefficient for $R$. Standard errors are computed via the delta method, propagating covariance across platform-specific coefficients.

\begin{table}[htbp]
  \centering
  \caption{Post Distribution by Platform}
  \label{tab:posts_by_platform}
  \begin{tabular}{lrr}
    \toprule
    \textbf{Platform} & \textbf{Posts} & \textbf{\%} \\
    \midrule
    Twitter & 2,000,000 & 34.91 \\
    Telegram & 1,736,327 & 30.31 \\
    TikTok & 1,539,771 & 26.88 \\
    Truth Social & 452,926 & 7.91 \\
    \midrule
    \textbf{Total} & \textbf{5,729,024} & \textbf{100.00} \\
    \bottomrule
  \end{tabular}
\end{table}

\begin{figure*}[h]
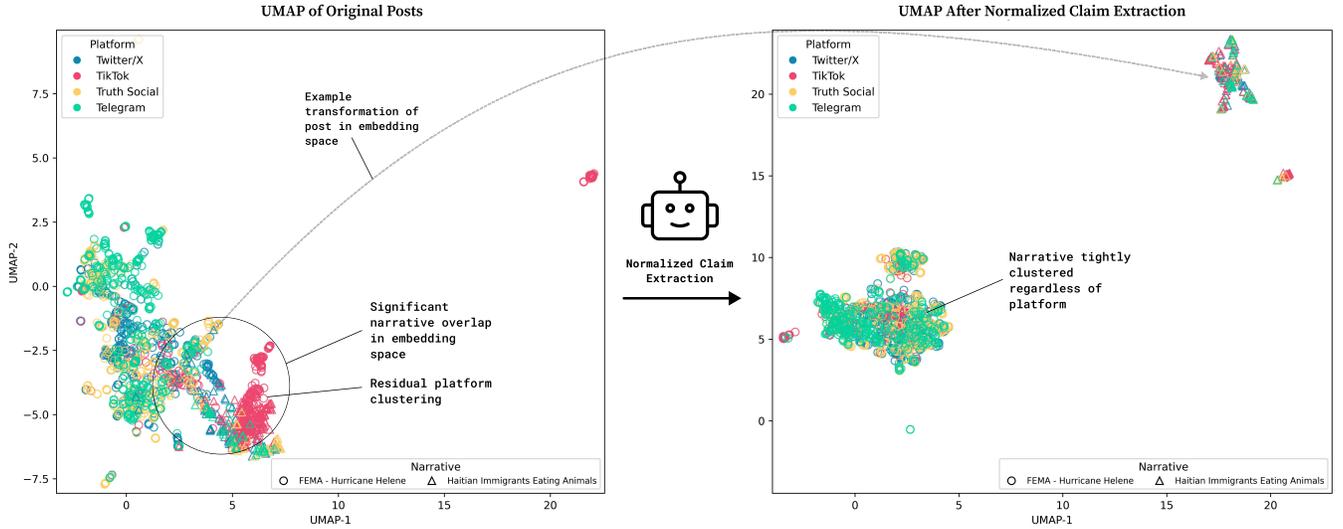
 
    \centering 
    \begin{overpic}[width=\textwidth]{figures/before-and-after-claims-v2.pdf}
    \end{overpic}

    \vspace{-1mm} 
\caption{UMAP projections of two separate narratives before (left) and after (right) applying our LLM-based claim extraction. Prior to normalization, key elements of the narratives are muddied across platforms, with content blending rather than separating cleanly. After claim extraction, coherent, separate narratives emerge, showing that platform-specific linguistic noise exerts less influence once the narrative is normalized.}
    \vspace{-3mm}
    \label{fig:umap-before-and-after-normalizer}
\end{figure*}

\section{Computational Complexity and Performance Analysis}
\label{app:complexity_analysis}

We analyze both theoretical complexity and empirical performance of the system on our deployment hardware.

\vspace{3pt}
\noindent
\textbf{Hardware Specifications}
\label{app:hardware_specs}

The full pipeline runs on heterogeneous hardware: an NVIDIA Quadro RTX 8000 (40GB VRAM) for both embedding generation and Gemma-4B-Instruct inference, and an Intel Xeon E5-2650 v4 @ 2.20GHz (30MB cache) for DP-Means clustering and discourse network updates.

\vspace{3pt}
\noindent
\textbf{Time Complexity}
\label{app:time_complexity}

Each component scales linearly with input size: claim extraction $O(nL)$ and embedding generation $O(nL)$, where $n$ is posts and $L$ is average sequence length; DP-Means clustering $O(nKd)$, where $K$ is current clusters and $d$ embedding dimension, with streaming updates avoiding $O(n^2)$ recomputation; and network construction $O(n \log n \cdot d)$ using FAISS-HNSW~\cite{malkov2018efficient} rather than $O(n^2d)$ exact search. Overall, the pipeline complexity is $O(n(L + Kd + \log n \cdot d))$, dominated by embedding and network construction.

\begin{table}[htbp] 
\caption{Component Performance Measurements}
\label{tab:component_performance}
\begin{tabular}{lr}
\toprule
\textbf{Component} & \textbf{Time Complexity} \\
\midrule
Claim Extraction (Gemma-4B) & $O(nL)$ \\
Embedding generation (MPNet) & $O(nL)$ \\
DP-Means clustering (streaming) & $O(nKd)$ \\
Network construction (FAISS-HNSW) & $O(n \log n \cdot d)$ \\
\midrule
\textbf{End-to-end pipeline} & $O(n(L + Kd + \log n \cdot d))$ \\
\bottomrule
\end{tabular}
\vspace{2pt}
\small
$n$: number of posts; $L$: average sequence length, $K$: current cluster count, $d$: embedding dimension
\end{table}




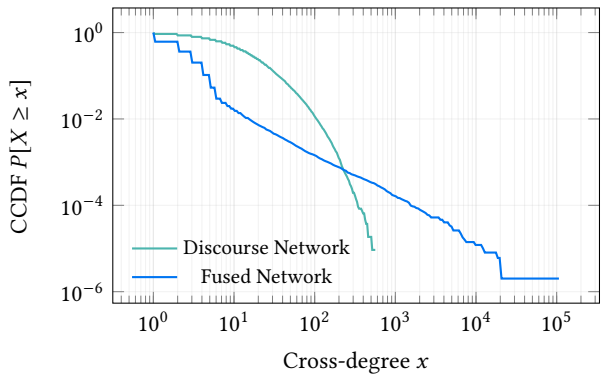
\begin{figure}[h]
  \begin{tikzpicture}
    \hspace{-12pt}
    \begin{axis}[
      width=0.95\linewidth,
      height=0.65\linewidth,
      xlabel={Cross-degree $x$},
      ylabel={CCDF $P[X \geq x]$},
      xmode=log,
      ymode=log,
      grid=both,
      grid style={opacity=0.2},
      legend style={at={(0.02,0.02)}, anchor=south west, draw=none, fill=none, font=\small},
    ]

    \addplot[Bridge, thick, mark=none]
        table[x=x,y=ccdf,col sep=comma]{csvs/ccdf_cap-based.csv};
      \addlegendentry{Discourse Network}

      \addplot[NoBridge, thick, mark=none]
        table[x=x,y=ccdf,col sep=comma]{csvs/ccdf_fusion.csv};
      \addlegendentry{Fused Network}

    \end{axis}
  \end{tikzpicture}
\caption{Complementary cumulative distribution function (CCDF) of cross-degree across networks. 
The discourse network shows substantially higher typical connectivity, with a median cross-degree of 9 and mean of 14.9 ($\sigma=21.2$), compared to the fused network where most users have $\leq 1$ cross-tie despite a few extreme hubs. 
The discourse network distribution thus reflects more widespread cross-platform ties among ordinary users, while the fused network is dominated by a small set of super-connectors.}
\label{fig:ccdf-cap}
  \label{fig:ccdf-cap}
\end{figure}

\begin{table*}[htbp]
\caption{\textbf{Example cross-platform narratives with early emergence.}
Each row lists a key narrative observed to spread predictably across platforms, emphasizing conspiratorial or procedural framing. Human-authored descriptions were generated using closest posts to centroid and narrative-level NPMI keywords.}
\small
\setlength{\tabcolsep}{6pt}
\renewcommand{\arraystretch}{1.4}
\begin{tabularx}{\linewidth}{X}
\toprule
\textbf{Narrative Description} \\
\midrule
\textbf{Weather control / geoengineering}: Claims that Hurricane Helene-related storms are artificially created by HAARP or NOAA. \\
\textbf{FEMA blocking or mishandling aid (Helene)}: Crisis-to-blame pivot in which hurricane footage becomes ``evidence'' of government obstruction. \\
\textbf{Pelosi pushing Biden off the ticket}: Elite-infighting narrative portraying Democratic leadership coup. \\
\textbf{``Jill Biden running the White House / Cabinet''}: Delegitimization through spousal-control trope. \\
\textbf{ActBlue unauthorized donations / fraud}: Financial-misconduct frame alleging illicit fundraising: ``fraud,'' ``illegal donors''. \\
\textbf{Wuhan lab + U.S. bioweapons collaboration}: Cross-national blame frame merging COVID and defense tropes. \\
\textbf{Project 2025 will create jobs}: Economic reframing of extremist policy blueprint as pragmatic job-creation plan; mainstreaming of radical agenda. \\
\textbf{Child sex-trafficking surge}: QAnon-adjacent revival of hidden-abuse networks. \\
\textbf{Harris staff exodus / toxic office}: Moral-legitimacy and leadership-competence attack. \\
\textbf{Illegal-immigrant crime surge}: Fear-based othering narrative linking emergence to violence and instability. \\
\textbf{Purge the government / deep state}: Revolutionary or cleansing frame invoking mass firings or removals. \\
\bottomrule
\end{tabularx}
\label{tab:notable_narratives_expanded}
\end{table*}

\section{Dataset}

\begin{table}[htbp]
\centering
\caption{Task 1: emergence detection label distribution}
\label{tab:task1_distribution}
\resizebox{.7\columnwidth}{!}{%
\begin{tabular}{lrr}
\toprule
\textbf{Time Horizon} & \textbf{\% Positive} & \textbf{\% Negative} \\
\midrule
3 days & 13.4\% & 86.6\% \\
7 days & 16.1\% & 83.9\% \\
14 days & 18.4\% & 81.6\% \\
\bottomrule
\end{tabular}
}
\end{table}

\section{Ablation}
\label{app:ablation}
\begin{table}[t]
\centering
\caption{Ablation study on 3-day emergence detection (AUC, mean $\pm$ std over 10 runs). Results show the contribution of individual feature groups and their combinations.}
\label{tab:ablation}
\resizebox{.8\columnwidth}{!}{%
\begin{tabular}{lc}
\toprule
\textbf{Feature Set} & \textbf{AUC} \\
\midrule
Total cross-platform connection count & $0.64_{\pm 0.01}$ \\
Active connection ratio & $0.73_{\pm 0.00}$ \\
Active connection count & $0.77_{\pm 0.01}$ \\
Active + Total connections & $0.88_{\pm 0.01}$ \\
\midrule
Features without claim extraction & $0.79_{\pm 0.01}$ \\
Features with claim extraction & $0.88_{\pm 0.01}$ \\
\bottomrule
\end{tabular}%
}
\end{table}

\clearpage

\end{document}